\RequirePackage{lineno}
\documentclass[twocolumn,showpacs,preprintnumbers,amsmath,amssymb,superscriptaddress,aps,floatfix]{revtex4}

\usepackage{longtable}
\usepackage{graphicx}
\usepackage{dcolumn}
\usepackage{bm}
\usepackage{color}
\usepackage{subfigure}
\usepackage{multirow}
\usepackage{amssymb}
\usepackage{mdwlist}
\usepackage{enumitem}
\usepackage{soul}

\def\today{\space\number\day\space\ifcase\month\or January\or February\or
    March\or April\or May\or June\or July\or August\or September\or October\or
    November\or December\fi\space\number\year}

\newcommand{\iso}[2]{$^{#1}$#2}

\newcommand{\rpmt}{R_{\mbox{{\tiny PMT}}}}

\begin{document}


\preprint{Draft 0.3}

\title{A Model for the Secondary Scintillation Pulse Shape from a Gas Proportional Scintillation Counter}

\newcommand{\llnl}{Lawrence Livermore National Laboratory, Livermore CA 94550}
\newcommand{\ucb}{University of California, Berkeley, CA, Currently at Lawrence Berkeley National Laboratory, Berkeley, CA}
\author{K.~Kazkaz}\affiliation{\llnl}
\author{T.\,H.\,Y.~Joshi}\affiliation{\llnl}\affiliation{\ucb}

\date{\today}

\begin{abstract}
Proportional scintillation counters (PSCs), both single- and dual-phase, can measure the scintillation (S1) and ionization (S2) channels from particle interactions within the detector volume.
The signal obtained from these detectors depends first on the physics of the medium (the initial scintillation and ionization), and second how the physics of the detector manipulates the resulting photons and liberated electrons.
In this paper we develop a model of the detector physics that incorporates event topology, detector geometry, electric field configuration, purity, optical properties of components, and wavelength shifters.
We present an analytic form of the model, which allows for general study of detector design and operation, and a Monte Carlo model which enables a more detailed exploration of S2 events.
This model may be used to study systematic effects in currents detectors such as energy and position reconstruction, pulse shape discrimination, event topology, dead time calculations, purity, and electric field uniformity.
We present a comparison of this model with experimental data collected with an argon gas proportional scintillation counter (GPSC), operated at 20 C and 1 bar, and irradiated with an internal, collimated \iso{55}{Fe} source.
Additionally we discuss how the model may be incorporated in Monte Carlo simulations of both GPSCs and dual-phase detectors, increasing the reliability of the simulation results and allowing for tests of the experimental data analysis algorithms.

\end{abstract}

\maketitle
%
%
%
%
%
\section{Introduction}
\label{s:Intro}

Proportional scintillation counters (PSCs) utilize electroluminescence to convert charge signals to scintillation signals, amplifying them in the process
The low thresholds of these systems have made single-phase, or gas proportional scintillation counters (GPSCs) particularly useful for the study of x-rays (see Ref.~\cite{Cov2004} and references therein). 
Dual-phase proportional scintillation counters also exhibit low thresholds, but with the increased target density afforded by the liquid target volume.  
Such detectors are a current technology of choice in direct searches for weakly interacting massive particles (WIMPs)~\cite{Ang2011, Aki2012a, Apr2012, Ale2013, Ake2014}, as well as in efforts for detecting coherent elastic neutrino-nucleus scatters (CENNS)~\cite{Hag2004, Aki2012b}.

\begin{figure}[b]
\centering
\includegraphics[width=8.5cm]{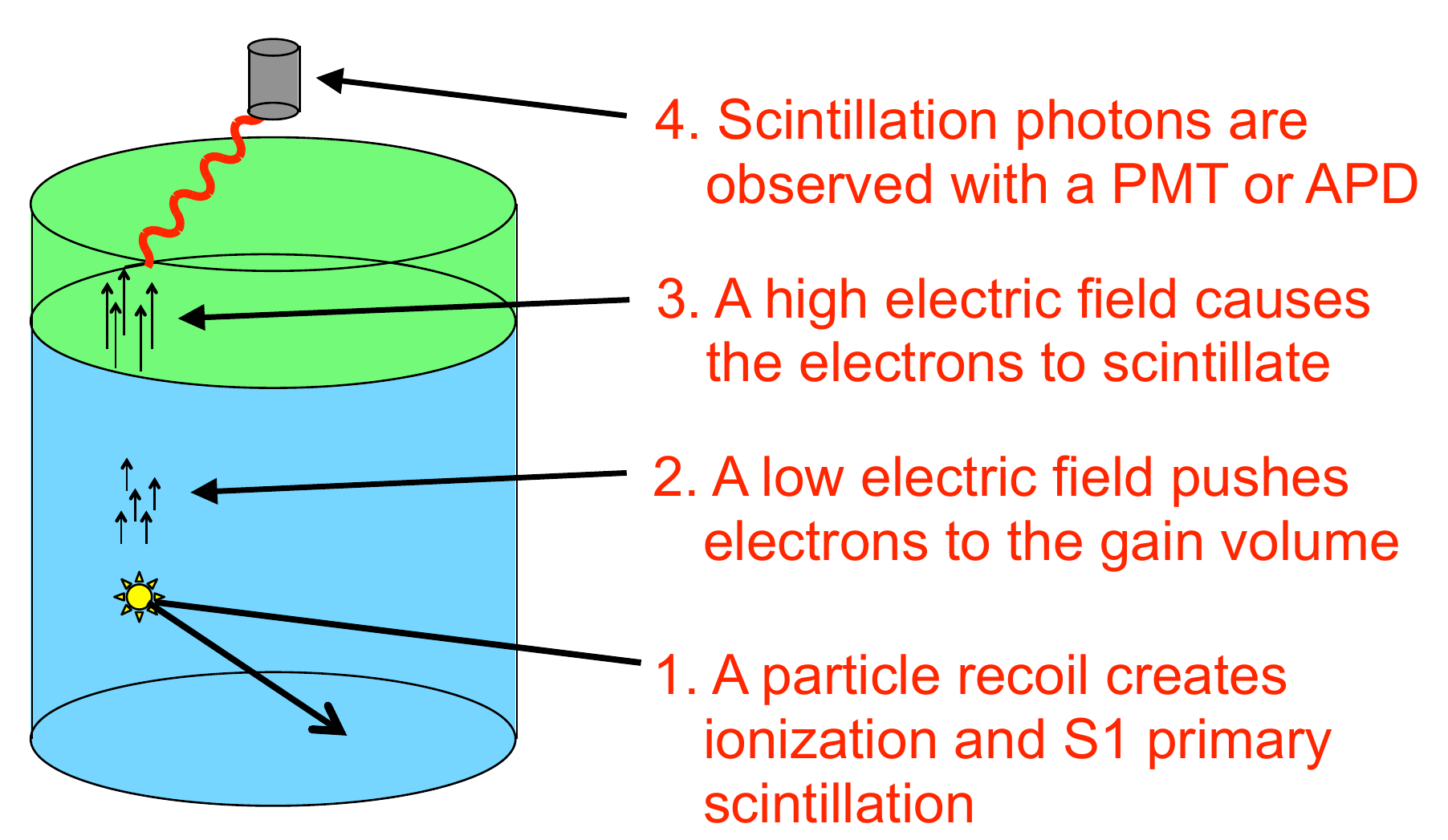}
\parbox{8.5cm}{\vspace{5pt}\caption{\small{Schematic of how the S2 signal is generated. In most dual-phase searches, one or two arrays of PMTs are used. Here, a single PMT is shown for illustrative purposes. The S1 light is emitted entirely in step 1 of the cartoon, and is also detected using PMTs or APDs.}}
\label{fig:DualPhaseCartoon}}
\end{figure}

GPSCs and dual-phase detectors operate in similar fashions. In both detector types, particle interaction energy loss mechanisms include scintillation and ionization. 
If the initial energy deposition is sufficiently high, the scintillation channel can be observed via a flash of prompt light (S1). 
Ionized electrons, however, are drifted via a low electric field to a gain volume with a high electric field. 
In noble gasses, these electrons create excited dimers during their transit through the gain volume, and the dimers subsequently relax via emission of secondary scintillation light (S2)~\cite{Krylov2002}, which is typically observed using a photomultiplier tube (PMT) or avalanche photodiode (APD).
A cartoon of the process is shown in Fig.~\ref{fig:DualPhaseCartoon}.

The electroluminescence, or S2, response of PSCs depends both on properties of the target medium, and the detector design and operating parameters.
For example, a liquid argon target at a given electric field, temperature, and pressure will have a consistent response to the same incident radiation, but the final S2 signal is affected by drift lengths, materials used, purity, and variations in the electric field between the target and electroluminescence region.
Understanding both the inherent properties of the target medium and the detector response is needed not only to correctly interpret experimental data, but also in understanding background events of unknown source or strength, as well as predicting performance of future detectors.

Unfortunately, the medium's response is obscured by the detector effects.
Therefore, each step in the chain of processes, from the initial particle interaction to the physical acquisition of data, must be understood individually to confidently understand the experimental data.
This is important because the S2 signal is used in calculations of event energy reconstruction, event position reconstruction, background discrimination, dead time calculations, event topology, dynamic energy range, and purity.
We have therefore developed a model of the chain of processes of so-called S2 pulses from PSCs.
This model is then tested against experimental data from an argon GPSC, with comparisons made to both the energy spectrum and the mean S2 event shape.

The time structure of the S2 signal depends on many factors, each of which can affect key features of the pulse shape. 
It is possible that a single, simple feature of the pulse shape can provide direct measurements of the properties of the S2 volume. 
For example, the triplet scintillation time constant is relatively straightforward to measure (Section~\ref{s:PulseShapeComp}), and provides information on impurity levels within the gaseous volume~\cite{Amsler2008}. 
Most of the features, however, are a result of an interplay between detector geometry, photon and electron amplification levels, singlet / triplet ratio, multiple scintillation decay times, electron drift speed and diffusion, reflectivity of detector components, and more. 
In addition to these physical effects, the algorithm used in analysis of the experimental data stream can also create systematic effects~\cite{Kazkaz2010}.

The majority of inputs to our model are obtained from experimental measurements in the literature. 
To the authors' knowledge the only parameters not provided in the literature are the singlet / triplet ratio of the S2 scintillation, which is generally considered to be zero, and the scattering length of scintillation light in the wavelength shifter used in our experimental GPSC. 
Once the full complement of model parameters of a detector are known, they can be used in a Monte Carlo calculation to determine the effects of changing a parameter, to study the causes of ambiguous signals, and to make reliable predictions of the performance of existing and future detectors.

In Section~\ref{s:AnalyticPulseShapeModel}, we introduce a simplified model and calculate the pulse shape analytically. We then corroborate the analytic model against a Monte Carlo calculation in Section~\ref{s:MCPulseShapeModel}, and extend the study to include all model effects. 
In Section~\ref{s:PulseShapeComp} we compare the full model to experimental data and discuss the results and further application of the model. 
We discuss what additional effects should be incorporated to extend the model to a dual-phase detector in Section~\ref{s:Extension}.

%
%
\section{The Analytic Pulse Shape Model}
\label{s:AnalyticPulseShapeModel}

The cartoon of the S1 and S2 signals shown in Fig.~\ref{fig:DualPhaseCartoon} is a simplified one. 
We will use this cartoon, though, to build the basic mathematics of the model. 
There are two reasons we start with this analytic approach. 
The first reason is to gain an intuitive understanding of various effects via the explicit equations and the curves they generate. 
The second reason is to validate the initial Monte Carlo calculations detailed in Section~\ref{s:MCPulseShapeModel}.

The analytic model incorporates a simplified detector geometry,  a constant electron drift speed across the gain volume, a constant longitudinal and lateral width of the charge cloud, the singlet and triplet decay time constants, and continuous transitions into and out of the gain volume.
These calculations are performed for readout by a single circular PMT.
These constraints will be relaxed in the next section when building the Monte Carlo model.

\subsection{Simple Analytic Model}
\label{ss:SimplifiedAnalyticModel}

\begin{figure}[t]
\centering
\includegraphics[width=8.5cm]{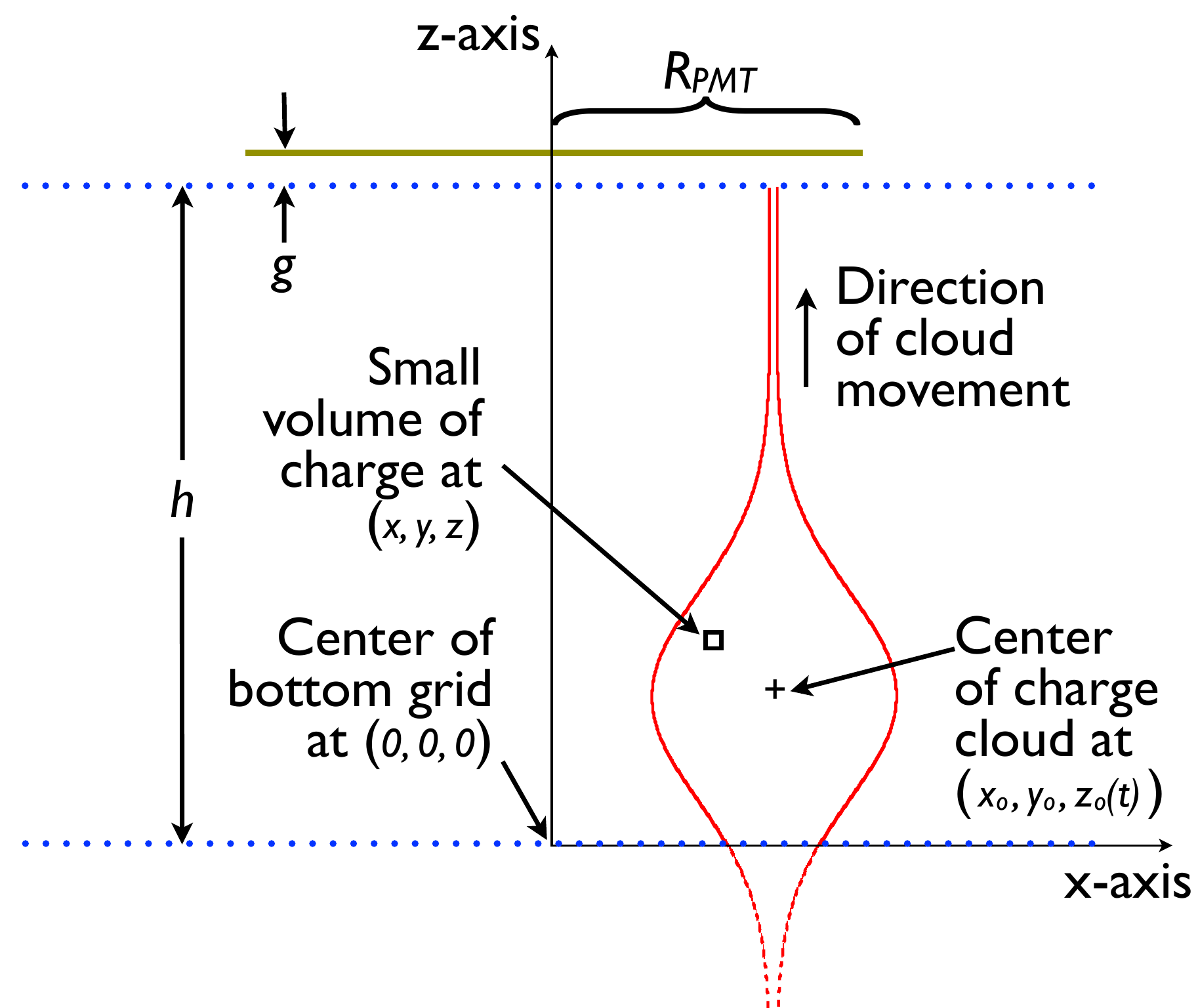}
\parbox{8.5cm}{\vspace{5pt}\caption{\small{Generalized geometry and coordinate system used in the mathematical formulation of the analytic pulse shape model. 
The dotted horizontal lines represent the grid wires that define the secondary volume. 
The solid horizontal line at the top of the figure is the window of the photomultiplier tube. 
$h$ is the height of the S2 volume, $g$ is the gap between the S2 volume and the window of the PMT, which itself has radius $\rpmt$. 
The charge cloud, as illustrated by the red curves, is a three-dimensional Gaussian with transverse width $\sigma_t$ and longitudinal width $\sigma_l$. 
The center of the cloud moves up through the S2 volume at velocity $v_2$. 
This figure is presented in two dimensions for clarity, although both the analytic and Monte Carlo calculations assume a three-dimensional geometry.}}
\label{fig:2DGaussianGeometry}}
\end{figure}

The basic geometrical setup is shown in Fig.~\ref{fig:2DGaussianGeometry}. 
From any given point $\left ( x, y, z \right )$, the geometric efficiency of a photon to hit the PMT window is given by 

\begin{widetext}
\begin{eqnarray}
\label{eq:Efficiency3D}
\eta( x, y, z ) = \int_0^{\rpmt} dr_p \int_0^{2\pi} d\theta_p  \frac{ \left ( h + g - z \right ) r_p}{4 \pi \left [ (\sqrt{x^2 + y^2} - r_p~cos\theta_p)^2 + (r_p~sin\theta_p)^2 + (h + g - z)^2 \right ]^{3/2}}
\end{eqnarray}

\begin{eqnarray}
\label{eq:GausChargeDensity}
\rho \left ( x, y, z, t \right ) = \frac{1}{2 \pi \sqrt{2 \pi}}\frac{1}{\sigma_t^2 \sigma_l} \mbox{~Exp} \left [ -\frac{1}{2} \left ( \frac{ \left ( x - x_0 \right )^2 + \left ( y - y_0 \right )^2}{\sigma_t^2} + \left( \frac{z - z_0 \left ( t \right ) }{\sigma_l} \right )^2 \right ) \right ] H \left ( 0, s_0 \right ) \times \left ( 1 - H \left ( h, s_h \right ) \right )
\end{eqnarray}
\end{widetext}

\noindent
where $r_p$ and $\theta_p$ are integration variables for the face of the PMT, and all other terms are defined in Fig.~\ref{fig:2DGaussianGeometry} and its caption. 
The distribution of charge within the drifting cloud is given by

\noindent
where $\sigma_t$ and $\sigma_l$ are defined in the caption of Fig.~\ref{fig:2DGaussianGeometry}. 
In Eq.~\eqref{eq:GausChargeDensity} the Heaviside function creates a more continuous cutoff in the cloud below $z = 0$ and above $z = h$:

\begin{equation}
\label{eq:Heaviside}
H \left ( z_{off}, \alpha \right ) = \frac{1}{1 + e^{-(z-z_{off})/\alpha}}
\end{equation}

\noindent
In Eq.~\eqref{eq:Heaviside}, the offset parameter $z_{off}$ is expressed in units of distance, although dividing by the macroscopic electron drift speed transforms the Heaviside function from the spatial to the temporal realm.
The variable $\alpha$ sets the strength of the cutoff, and is used to represent the sharpness of the electric field gradient across the grid wires that define the S2 volume.
The best choice for $\alpha$ will depend on the grids or surfaces that define the boundaries of the S2 volume.
In these simplified calculation $\alpha$ is set to 0.2~mm.
For these calculations similar electric field transitions are assumed across the two boundaries. 
Thus in Eq.~\eqref{eq:GausChargeDensity} $s_0 = s_h$.

The center of the charge cloud ($z_0$) in Fig.~\ref{fig:2DGaussianGeometry} drifts upward with characteristic velocity $v_2$, and may therefore defined as,

\begin{equation}
\label{eq:ZOfT}
z_0 \left ( t \right ) = z_0(0) + v_2 \, t
\end{equation}

\noindent
where $z_0(0)$ is the point at which the charge cloud is centered at the begining of our calculation.
To accurately represent the leading edge of the event signal, integration of the charge cloud drift across the gain volume is initialized with the charge cloud centroid $-3 \sigma_l$ below the gain volume.

Just as the center of the charge cloud drifts as a function of time, so to do $\sigma_l$ and $\sigma_t$. $\sigma_l$ evolves according to the equation
\begin{equation}
\label{eq:GausWidth}
\sigma_{l(t)} = \sqrt{C_{D,l(t)} ~ \Delta t}
\end{equation}

\noindent
where $C_{D,l(t)}$ is the constant of longitudinal (transverse) diffusion. 
Incorporating time dependence of $\sigma_l$ and $\sigma_t$ made numerical calculations computationally intensive, reducing the utility of this model.
These terms are therefore held constant, and selected typical values calculated based on the experimental setup.
Time-dependence of these parameters is re-introduced in the Monte Carlo calculations discussed in Section~\ref{ss:MonteCarloModel}.

The finite liftetime of the scintillation mechanism is also included.
In the case of noble elements, scintillation light is emitted during the relaxation of eximers, which can be produced in a singlet or triplet state~\cite{Suzuki1979}.
Singlet eximers relax quickly relative to most PMT response times, while the slower triplet state extends the pulse beyond the time that the last ionization electron exits the gain volume. 
To incorporate the effect of finite singlet and triplet eximer lifetimes, we convolve the light emission time with two exponentials, leading to the equation

\begin{widetext}
\begin{equation}
\label{eq:ModifiedShape}
S_{\mbox{s/t}} \left ( x, y, z, t \right ) = \int_0^t \eta(x,y,z) ~ \rho(x, y, z, t') ~ \left ( \frac{F_s}{\tau_s} e^{-(t-t')/\tau_s} + \frac{1 - F_s}{\tau_t} e^{-(t-t')/\tau_t} \right ) dt'
\end{equation}
\end{widetext}

\noindent
where $\tau_s$ and $\tau_t$ are the lifetime of singlet and triplet eximers, respectively. 
$F_s$ is the fraction of singlet states, making $1 - F_s$ the fraction of triplet states. 
$F_s$ is a free parameter in our model that is not, to our knowledge, corroborated in the existing literature.

The final step in the analytic treatment is to perform the spatial integration:

\begin{equation}
\label{eq:3DFinalShape}
S\left(t\right) = \int_{-\infty}^\infty dx \int_{-\infty}^\infty dy \int_{-\infty}^\infty dz ~S_{\mbox{s/t}} \left ( x, y, z, t \right )
\end{equation}

\subsection{Application of Analytic Model}
\label{ss:AppliedAnalyticModel}

\begin{table*}[t]
\caption{\small{Parameters used in the analytic calculations of the pulse shape.}}
\begin{center}
\begin{tabular}{|l|l|l|}
\hline
Parameter & Value & Source \\
\hline
Height of S2 volume ($h$) & 38.735 mm & $\begin{array}{l} \mbox{Measurement of physical detector} \\ \mbox{(See Section~\ref{ss:MCValidation})} \end{array}$ \\
Gap between S2 volume and PMT ($g$) & 1 mm & $\begin{array}{l} \mbox{Measurement of physical detector} \\ \mbox{(See Section~\ref{ss:MCValidation})} \end{array}$ \\
Radius of PMT ($\rpmt$) & 23 mm & Hamamatsu specification sheet~\cite{Hamamatsu6522}\\
Initial cloud center ($x_o, y_o, z_o(0)$) & (0, 0, -5.7 mm) & Given starting value (See Section~\ref{ss:SimplifiedAnalyticModel})\\
Drift velocity in S2 volume ($v_2$) & 6.378 mm/$\mu$s & {\it magboltz} calculation, v10.6~\cite{Biagi1999}\\
Heaviside transition strength ($s_0$) & 0.2 mm & See Section~\ref{ss:SimplifiedAnalyticModel} \\
Singlet fraction ($F_s$) & 0, 0.5, 1 & Free parameter \\
Singlet lifetime ($\tau_s$) & 5 ns & Ref.~\cite{Suzuki1979} \\
Triplet lifetime ($\tau_t$) & 2.7 $\mu$s\footnote{See Section~\ref{s:PulseShapeComp} for a discussion of the choice of $\tau_t$.} & Ref.~\cite{Amsler2008}, Ref.~\cite{Suzuki1979} \\
Longitudinal charge cloud width ($\sigma_l$) & 1.9 mm & Typical value (see Section~\ref{sss:NonConstantSlSt}) \\
Transverse charge cloud width ($\sigma_t$) & 4.5 mm & Typical value (see Section~\ref{sss:NonConstantSlSt}) \\
\hline
\end{tabular}
\end{center}
\label{tab:AnalyticParameters}
\end{table*}

Eq.~\eqref{eq:3DFinalShape} was numerically integrated over to obtain three conceptual pulse shapes, each with different values of $F_s$ (Fig.~\ref{fig:MCAnalyticComparison}). 
To avoid instabilities in the calculation, time resolution finer than the singlet scintillation decay time of 5~ns was needed, 1~ns was used. 
This temporal resolution, combined with five integration variables ($r_p$ and $\theta_p$ from Eq.~\eqref{eq:Efficiency3D} and $x$, $y$, and $z$ from Eq.~\eqref{eq:3DFinalShape}) resulted in a computational cost of 400~CPU-hours and required post processing to eliminate artifacts from the numerical integration. 
At this point the analytic model contains sufficient complexity to serve as an effective validation tool for Monte Carlo simulations. 
Such simulations are are computationally efficient and enable the introduction of additional physical processes that further impact the response of proportional scintillation counters. 

%
%
\section{The Monte Carlo Pulse Shape Model}
\label{s:MCPulseShapeModel}

\subsection{Validation of the Monte Carlo Calculations}
\label{ss:MCValidation}

A simulation geometry consistent with that outlined in Section~\ref{ss:SimplifiedAnalyticModel} was created using LUXSim~\cite{Akerib2011}, a Geant4-based~\cite{Agostinelli2003,Allison2006} simulation package (Geant4 version 4.9.4.p04). 
Additionally the physical processes considered in this initial simulation were set to be consistent with Section~\ref{ss:SimplifiedAnalyticModel}.

For the Monte Carlo calculations, each simulated event incorporated a number of electrons with Poisson mean of 230, roughly matching the number of ionization electrons from a 6-keV X-ray given a 26.4-eV work function~\cite{Jes1955}. 
These ionization electrons were then distributed with a probability given by Eq.~\eqref{eq:GausChargeDensity}, assuming $t = 0$ and given the constant values of $\sigma_l$ and $\sigma_t$ shown in Table~\ref{tab:AnalyticParameters}. 
Each electron was allowed to drift only parallel to the z-axis, with no additional diffusion. 
Because Geant4 does not track electrons below 250~eV, we employed user code to track the location of the electrons as they drifted through both the S1 and S2 volumes at a constant vertical velocity of $v_2$. 
Optical photons were then generated in the S2 volume at the changing electron locations, and stochastically increased the photons' times of emission according to the singlet and triplet decays times, weighted by the choice of $F_s$. 
Within the physical detector, we had a TPB-coated slide 1~mm above the S2 volume, and for the purposes of this initial Monte Carlo calculation, this slide took place of the PMT window shown in Fig.~\ref{fig:2DGaussianGeometry}.

\begin{figure}[b]
\centering
\includegraphics[width=8.5cm]{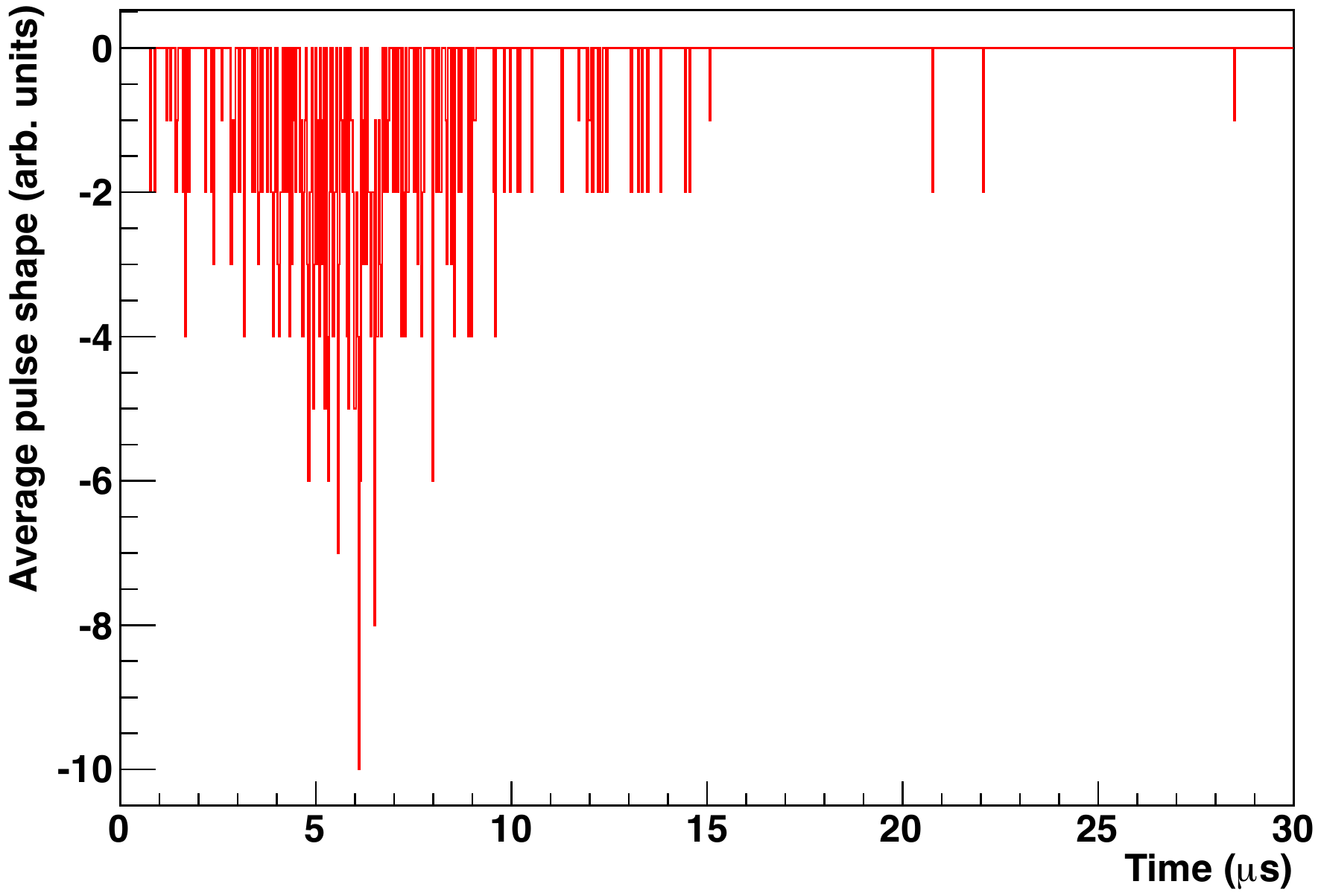}
\parbox{8.5cm}{\vspace{5pt}\caption{\small{A single simulated train of photons from ionization electrons drifting through the S2 volume. For this trace, we used $F_s$ = 0.5.}}
\label{fig:MCAnalyticComparison_SingleTrace}}
\end{figure}

\begin{figure}[t]
\centering
\includegraphics[width=8.5cm]{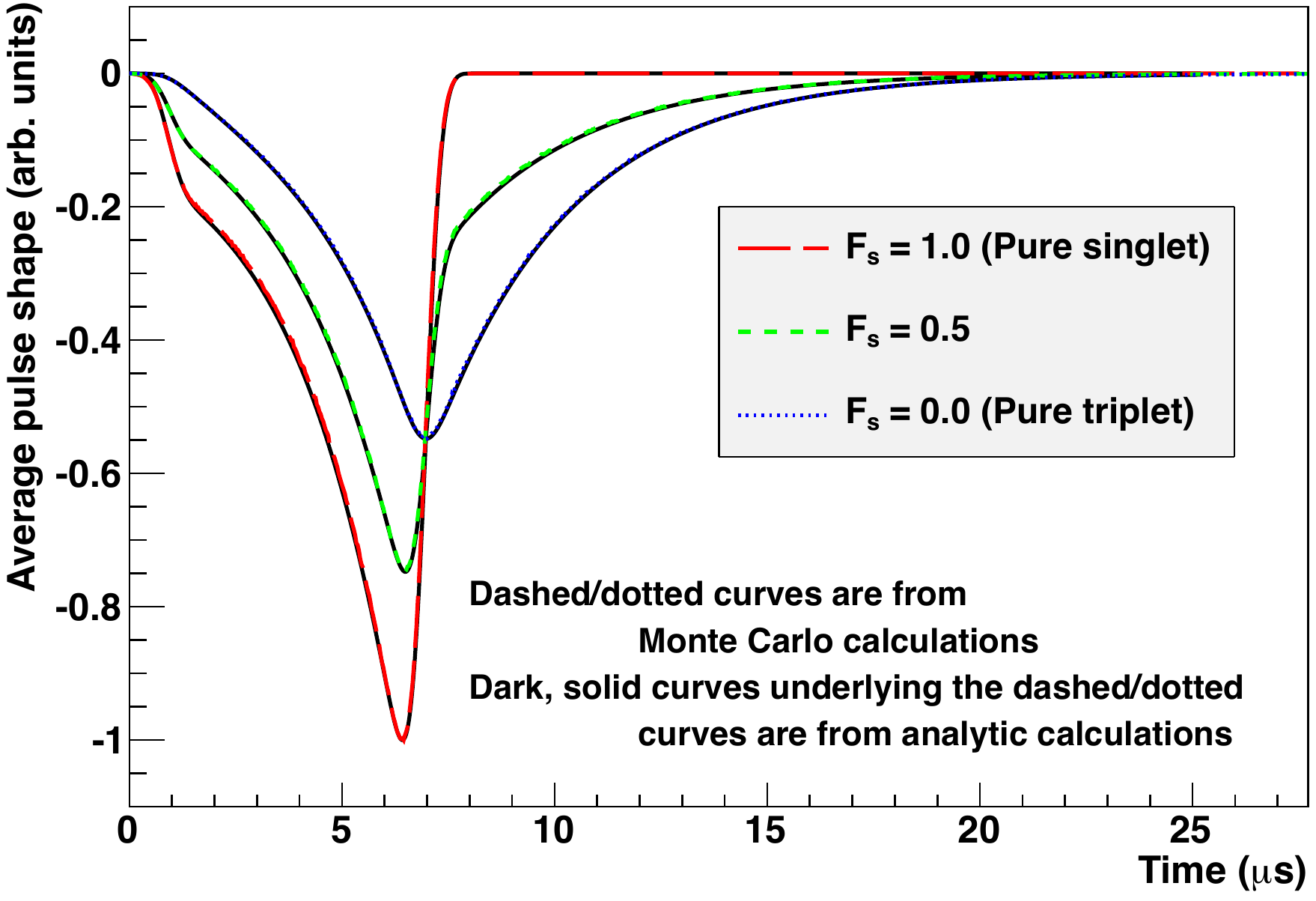}
\parbox{8.5cm}{\vspace{5pt}\caption{\small{Comparison of the Monte Carlo and analytic calculations. The dashed curves of the Monte Carlo calculations are plotted on top of the solid curves of the analytic calculations. There is no appreciable difference between the analytic and Monte Carlo results.}}
\label{fig:MCAnalyticComparison}}
\end{figure}

Performing this Monte Carlo calculation resulted in a time-varying train of photons incident on the TPB-coated slide (Fig.~\ref{fig:MCAnalyticComparison_SingleTrace}). 
100,000 individual traces were calculated and summed to obtain an average pulse shape. 
Figure~\ref{fig:MCAnalyticComparison} shows a comparison between our Monte Carlo and analytic calculations. 
The close agreement between the analytic and Monte Carlo curves gave confidence that the physics considered in the analytic model was being correctly handled in the Monte Carlo simulation.
The simulation required approximately 3 CPU-hours to run, and provided results indistinguishable from the analytic calculation. 

\subsection{Extending the Monte Carlo Calculations}
\label{ss:MonteCarloModel}

The Monte Carlo simulation was extended to include additional physical phenomena that impact the signals from gas proportional scintillation counters.  
Table~\ref{tab:AnalyticParameters} lists precise values and their sources that are used in the extended Monte Carlo simulations. 
The remainder of this section is used to describe effects that were added to the extended simulation.

\subsubsection{Smoothly-varying electric field}

\begin{figure}[t]
\centering
\includegraphics[width=8.5cm]{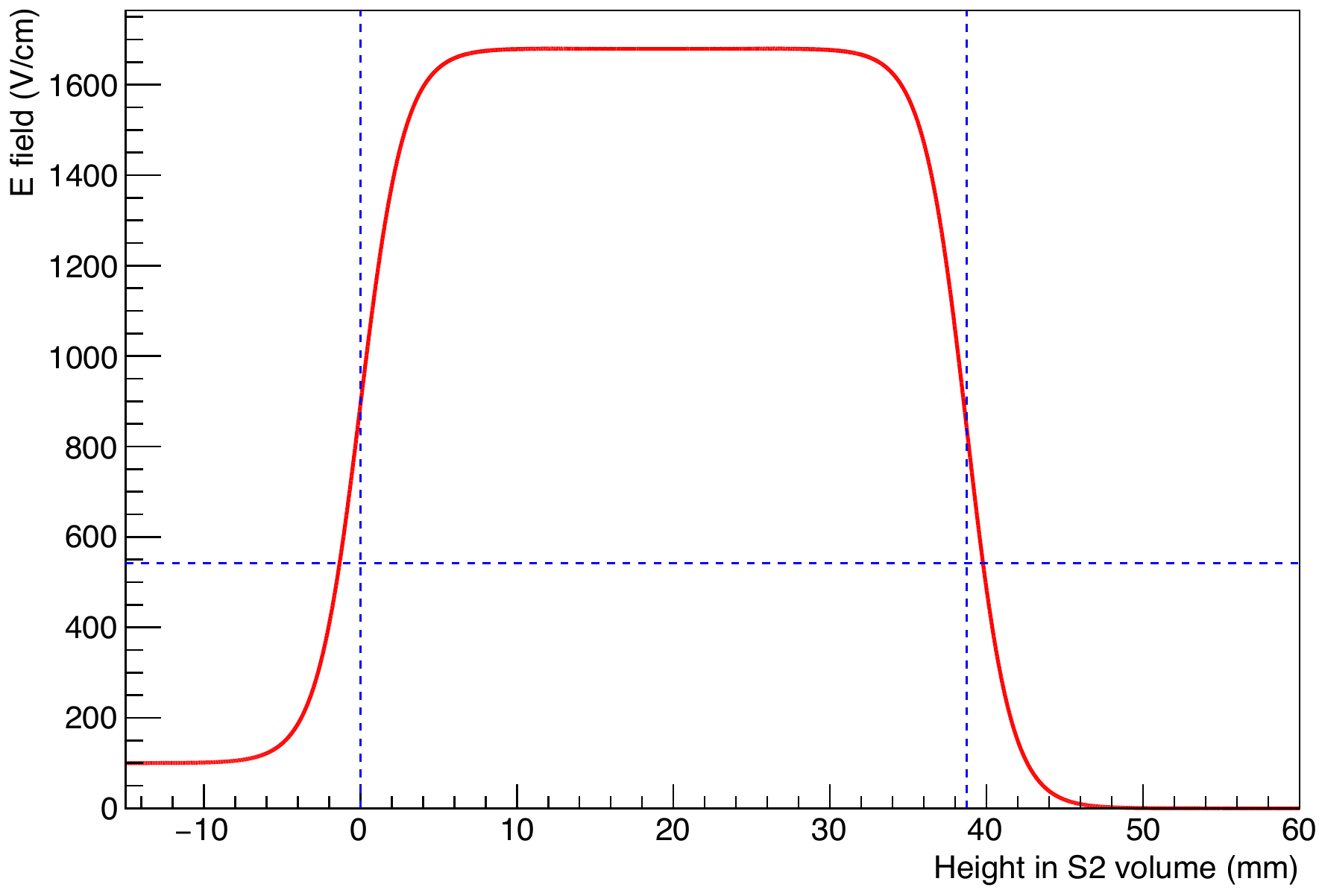}
\parbox{8.5cm}{\vspace{5pt}\caption{\small{Electric field profile for the single-phase detector. We assume a simple model for the electric field that varies only with the height, and not with the transverse dimensions. The dotted vertical lines shows the extent of the S2 volume. The horizontal dotted line shows the threshold for secondary scintillation (see Section~\ref{sss:FanoPhoton}), and extended slightly outside the bounds of the S2 volume.}}
\label{fig:EField}}
\end{figure}

The electric field has a direct effect on the average electron drift speed, longitudinal and transverse diffusion constants, photon production, and electron amplification. 
A continuous one-dimensional model of the electric field was used.
The field model, Fig.~\ref{fig:EField}, smoothly transitions from 100~V/cm in the bulk of the S1 volume, to 1680~V/cm in the bulk of the S2 volume, to 0~V/cm above the S2 volume.

\subsubsection{Multiple drift speeds, variable $\sigma_l$ and $\sigma_t$}
\label{sss:NonConstantSlSt}

\begin{figure}[t]
\centering
\includegraphics[width=8.5cm]{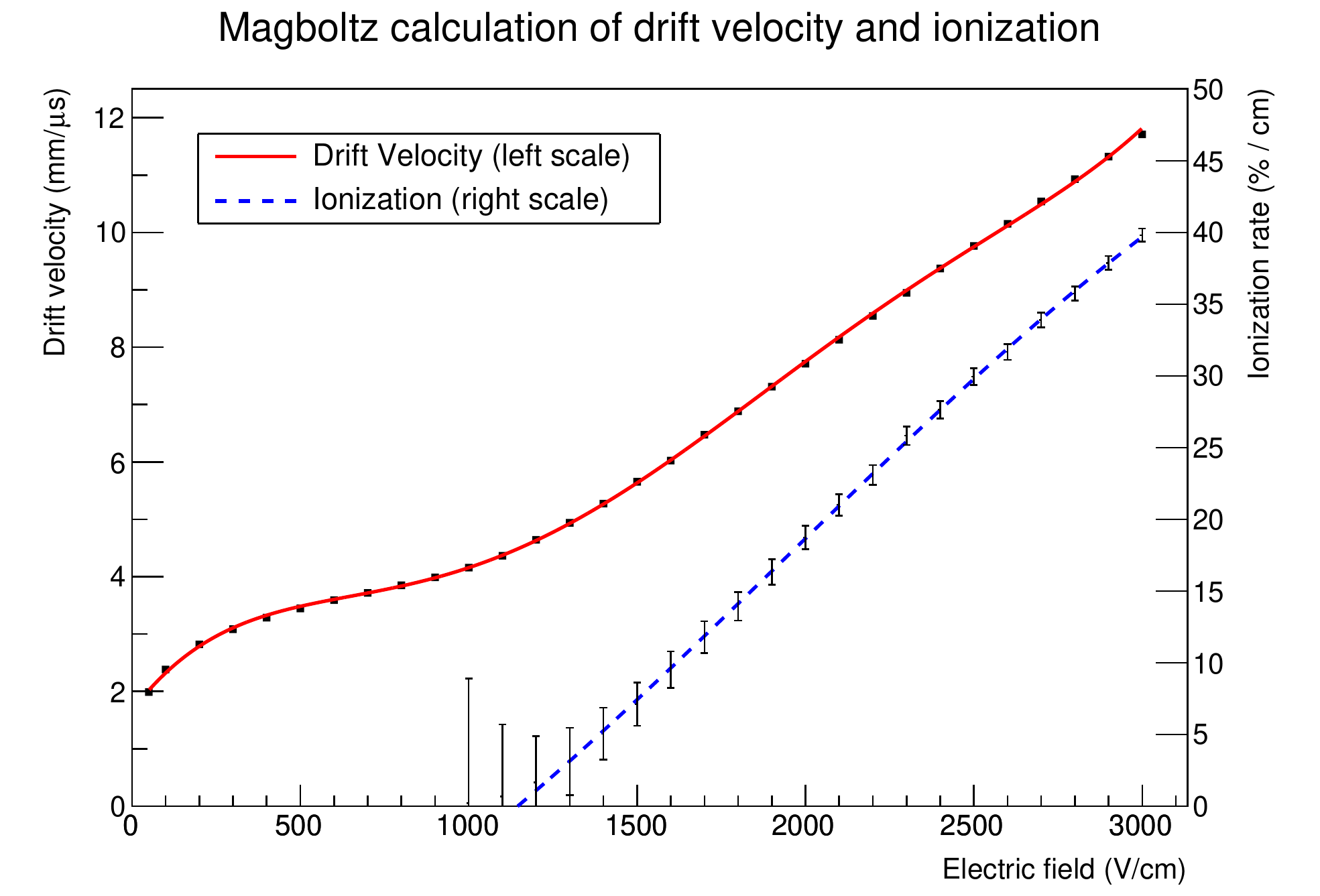}
\parbox{8.5cm}{\vspace{5pt}\caption{\small{Average drift speed and ionization rate calculated using $magboltz$~\cite{Biagi1999}. We applied an empirical fit between the data points to provide smooth interpolation between data points.}}
\label{fig:DriftSpeed}}
\end{figure}

The average electron drift speed depends in part on the electric field. 
$Magboltz$~\cite{Biagi1999} was used to calculate the drift speed as a function of electric field (Fig.~\ref{fig:DriftSpeed}), while keeping the pressure, temperature, and composition constant.
These values were used in Monte Carlo calculations. 

\begin{figure}[t]
\centering
\includegraphics[width=8.5cm]{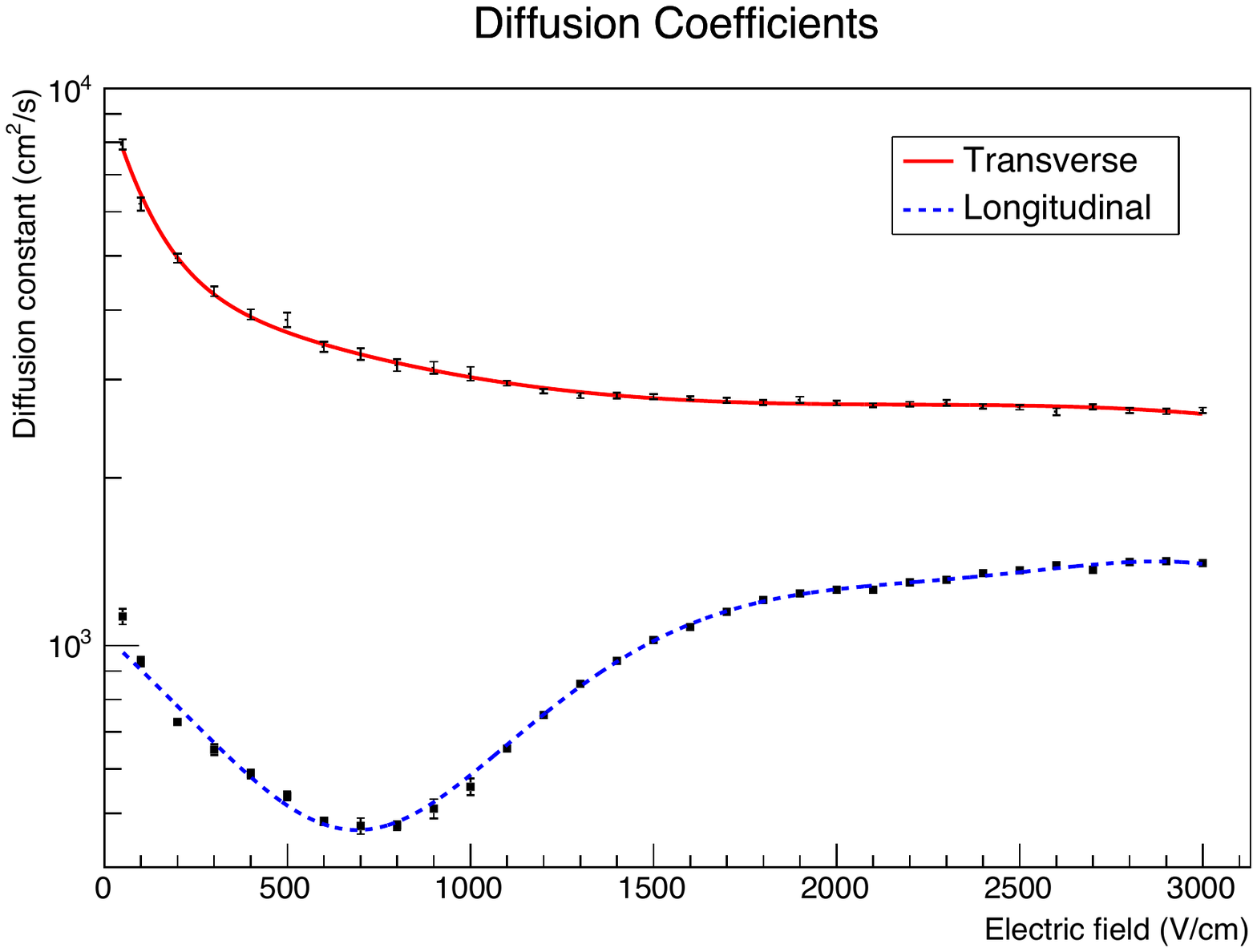}
\parbox{8.5cm}{\vspace{5pt}\caption{\small{Transverse and lateral diffusion constants. As with the drift speed, we used $magboltz$~\cite{Biagi1999} to calculate the dependency on the electric field. Also as with the drift speed, we employed empirical fits to the data points to provide smooth interpolation.}}
\label{fig:DiffusionConstants}}
\end{figure}

The dispersion of the drifting charge cloud is expected to vary with time, Eq.~\eqref{eq:GausWidth}.
Additionally, the diffusion constants $C_{D,l}$ and $C_{D,t}$ are functions of electric field (Fig.~\ref{fig:DiffusionConstants}), and therefore vary across the S1 and to the S2 volume boundaries. 
The $Magboltz$ code was used to calculated values for these terms, and the diffusion of charge in the MC simulation follows Eq.~\eqref{eq:GausWidth}.

\subsubsection{The location of energy depositions}

The location of energy deposition within a detector may impact the event shape in several ways: the depth of interaction dictates the amount of diffusion experienced by the charge cloud \cite{Sorensen2011}, the transverse position will impact the efficiency for collecting S2 light, and multi-site interactions (e.g. Compton scatters or photoelectric interaction followed by X-ray emission) result in spatially distinct charge clouds.  
The \iso{55}{Fe} calibration source used in the experimental system was collimated to provide X-ray events close to the detector axis. 
The geometry of this collimator was included in the simulation, and events produced using the Geant4 Radioactive Decay Manager, which generated both 5.9-keV and 6.4-keV X-rays from the electron-capture decay of \iso{55}{Fe}.

\subsubsection{Track length}

Depending upon the energy and type of ionizing particle, the spatial extent of the charge cloud at $t=0$ may not be point-like.
In the GPSC under study in this work, a 6-keV electron would leave a track length of approximately 1~mm~\cite{Berger1993}. 
With a maximum drift time of 0.4~$\mu$s, this effect may be discernible in our detector. 
Within the simulation, each step of energy deposition was divided by the work function to generate the number of ionization electrons in that step, and then track them as described above.

\subsubsection{Wavelength shifting}

The scintillation wavelength of argon, 128~nm, is too short to be observed by PMTs. 
S2 light must therefore be wavelength shifted to create an observable PMT signal.
A slide of glass with a thin layer of Tetraphenyl Butadine (TPB) on it, which not only shifted the wavelength of the photon, but randomized the angle of emission, was experimentally employed. 
The TPB wavelength shifter was implemented in Geant4 using the G4OpWLS class. 
The TPB absorption and emission spectra from \cite{Berlman1971} are used, and a relaxation time constant of 2~ns~\cite{Gryczynski1993} was used to describe this process. 
The areal density of the physical TPB coating was measured to be 0.062~mg/cm$^2$, corresponding to a thickness of 0.57461~$\mu$m. 
Because G4OpWLS is a bulk process, as opposed to occurring on the boundary between two materials, there was some concern over whether this sub-micron thickness was too narrow to allow for proper wavelength shifting and subsequent optical photon interaction ~\cite{Gumplinger2011}. 
Upon further analysis, it was determined that the narrow width of the TPB was not problematic.

\subsubsection{Photon generation and TPB scattering length}
\label{sss:PhotonsTPBScattering}

The integrated PMT pulse area of the experimental data was normalized by the area of a single photoelectron to obtain data in units of photoelectrons per event. 
This allowed for immediate comparison between experimental and simulation data. 
The PMT used in the experimental data was a Hamamatsu type R329-02 with UV glass, and over the range of the TPB emission spectrum the average quantum efficiency was 17\%.
A wavelength-dependent binomial process was included to describe the QE of the PMT in the simulation.

Given the pressure, temperature, and electric field in the experimental apparatus, a photon emission rate of 9.86 photons / electron / cm with a threshold of 542~V/cm is expected based on the results of ~\cite{Mon2008}.
Photon scattering in the TPB layer was described using a scattering length of $106.0 \pm 0.5$~nm. 
Reducing the scattering length has the same effect as increasing the reflectivity of the TPB.

Incorporating an appropriate attenuation length in the TPB layer would have a similar effect on the transmissivity of the wavelength-shifted light. 
But without a photodetector on both sides of the TPB layer, this degeneracy cannot be broken. 
It is therefore appropriate to consider the scattering length reported here as an effective scattering length, rather than a measurement of a physical scattering length. 

\subsubsection{Fano factor and secondary ionization}
\label{sss:FanoPhoton}

A constant Fano factor of 0.30 was applied to the creation of the ionization electrons to match that measured in~\cite{Mon2001}, though this is a simplification because both the charge yield~\cite{Szy2011} and the Fano factor itself~\cite{Len2014} vary with energy.

Additionally, using $magboltz$, we calculate a value of 1145\,V/cm for the threshold for secondary ionization in the experimental apparatus.
This value is lower than those suggested by ~\cite{Dia1986} and \cite{Mon2001}, which were considered during the experimental data collection campaign, however we find good agreement using the $magboltz$ derived value.
The operation of the S2 volume above the onset for secondary ionization resulted in broadening of the of the spectral peaks.
Probabilistic secondary ionization was therefore included in the MC model, using values calculated with $magboltz$.

One possible explanation for the differences in ionization thresholds might be found in the relatively short S2 volume used by in both ~\cite{Dia1986} and \cite{Mon2001}. 
With a short S2 volume, a very low ionization rate might not have been distinguishable from systematic fluctuations in the recombination of the primary energy deposition, or photon production rates in the S2 volume.

%
%
\section{Comparison Between Energy Spectra and Pulse Shapes, and Discussion}
\label{s:PulseShapeComp}

The MC model was tested against experimental data from a single-phase, room-temperature argon GPSC. 
The operation of this GPSC, as well as the experimental analysis algorithm, is detailed in Ref.~\cite{Kazkaz2010}. 
There are, however, important differences between Ref.~\cite{Kazkaz2010} and the current work. 
First, the current work uses pure argon gas and a window in front of the photomultiplier tube with a coating of tetraphenyl butadiene (TPB) to shift the argon scintillation light from 128~nm to the 420-nm range~\cite{Lally1996}. 
Second, the pressure of the argon gas in the current work was 755 Torr at 21 $^\circ$C, and the electric fields were 100 V/cm in the S1 volume, and 1680 V/cm in the S2 volume. 
These voltages were used because they were the highest attainable at the given pressure that resulted in stable operation. 
Third, the gate time used in the current analysis was 1~$\mu$s, because of the longer scintillation decay time of pure argon versus argon with a small admixture of nitrogen~\cite{Amsler2008}.

\begin{figure}[b]
\centering
\includegraphics[width=8.5cm]{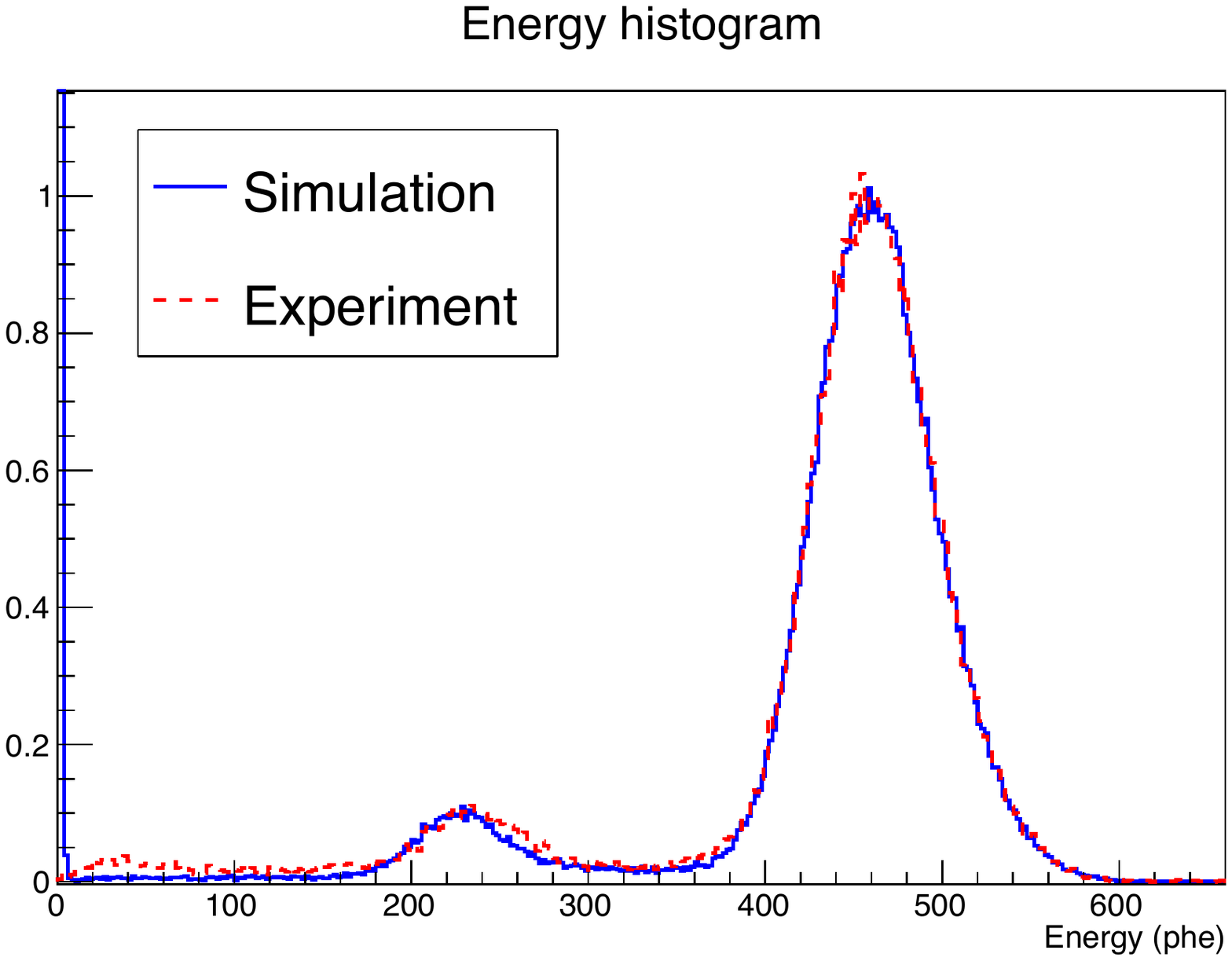}
\parbox{8.5cm}{\vspace{5pt}\caption{\small{Comparison of experimental and simulation energy spectra. The main peak near 460~photoelectrons comes from a dominant X-ray at 5.9~keV and a lower-intensity X-ray at 6.4 keV, while the smaller peak near 230~photoelectrons is a combination of four escape peaks resulting from K-L2, K-L3, K-M2, and K-M3 transitions. See~\cite{Kazkaz2010} for details.}}
\label{fig:EnergyComp}}
 \end{figure}

Figure~\ref{fig:EnergyComp} shows a comparison between simulated and experimental results, with strong agreement between calculated and experimental spectra given the parameters used. 

\begin{figure}[b]
\centering
\includegraphics[width=8.5cm]{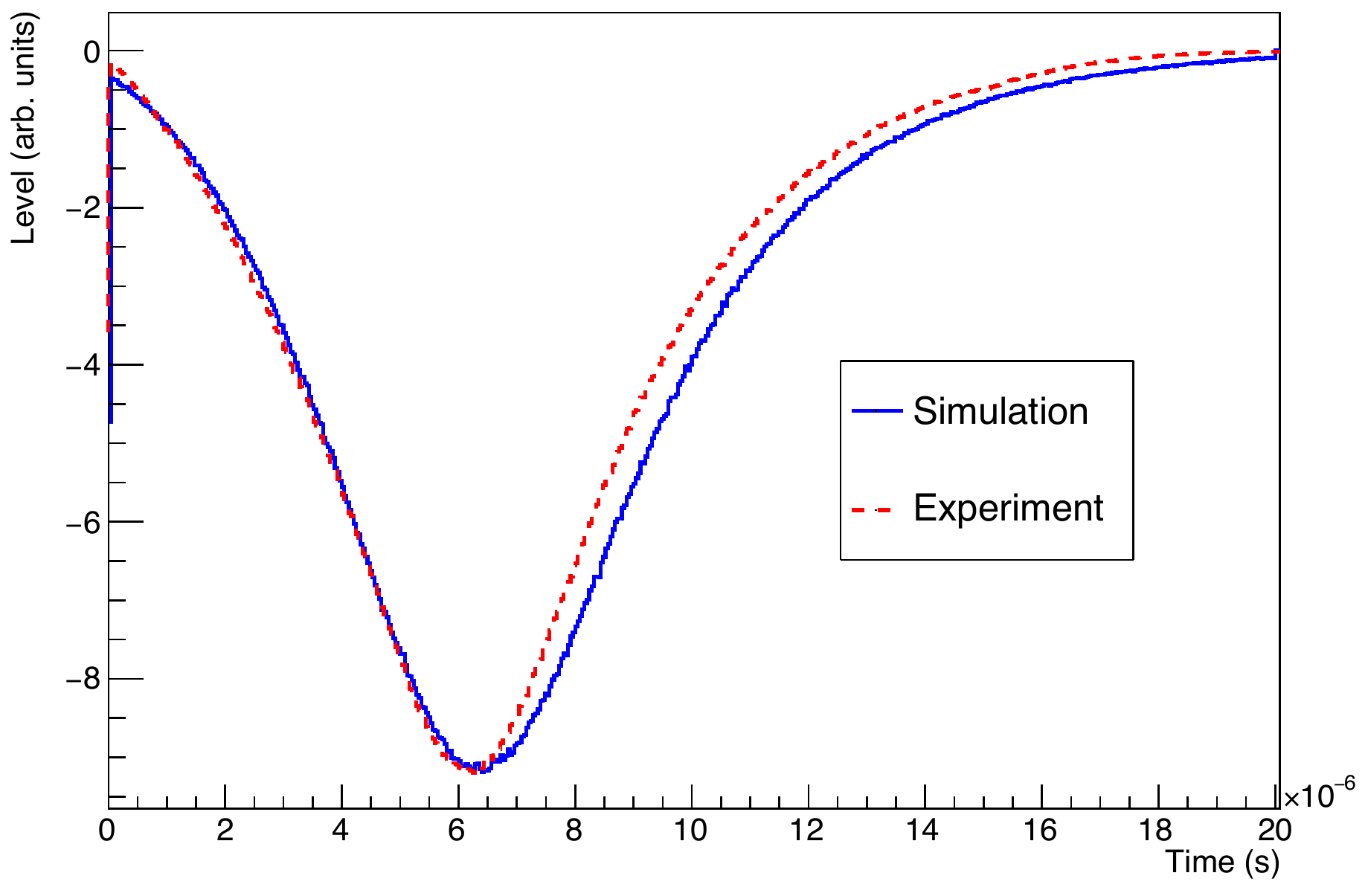}
\parbox{8.5cm}{\vspace{5pt}\caption{\small{Average shape of the S2 pulse. While the calculated and experimental curves match in the rising edge, the trailing edge is somewhat prolonged in the calculated curve. We attribute this to the simplified model of our electric field.}}
\label{fig:PulseShape}}
 \end{figure}

The close agreement between the experimental and simulated energy spectra suggests accurate handling of several interconnected physical processes.
The average event shape allows for additional comparisons between the simulated and experimental data.
The effects incorporated in Section~\ref{ss:MonteCarloModel} combined to give the calculated pulse shown in Fig.~\ref{fig:PulseShape}. 
The large-scale shape stems from the drifting of the ionization electrons through the S2 volume, combined with the $\mu$s-scale decay time of the triplet lifetime.

As can be seen from Fig.~\ref{fig:MCAnalyticComparison}, the lower the value of $F_s$, the slower the initial rise of the pulse. 
In Fig.~\ref{fig:PulseShape}, the y-intercept of the calculated pulse is greater than that of the experimental pulse.
Increasing $F_s$ would make this deviation even larger. 
We therefore set $F_s = 0$ when calculating the pulse shape, and given this limiting case, we conclude that singlet production in the S2 volume is negligibly small. 

Fitting an exponential decay to the experimental curve between 9 and 13~$\mu$s gives a decay constant of 2.76~$\mu$s, implying a partial nitrogen pressure of $\sim7.5\times10^{-6}$ Torr~\cite{Amsler2008}, or approximately one part in $10^8$. 
The purifier used to clean the argon gas (SAES model MC1500-903) did not remove nitrogen, so this value is considered to be an upper bound on any impurities within the argon gas.

The most noticeable mismatch between the experimental and calculated curves is in the trailing edge. 
Better agreement was sought by letting various parameters change from those obtained in the literature. 
The resulting curves, however, suffered from other inconsistencies with the experimental data. 

Ultimately, it was decided to forego modification to the independent physical parameters in the model. 
Not only were we unable to obtain a calculated curve that matched the entirety of the experimental curve, but those modifications were not physically well-motivated.

Given the attempts to obtain better agreement, we conclude that the input most likely to cause deviation from experimental results was the simple electric field model employed (Fig.~\ref{fig:EField}). 
A more accurate field calculation may resolve the discrepancy between the curves. 
A more careful treatment of the electric field might also reveal a large increase in the field near the top wires of the S2 volume, resulting in both increased luminescence and electron multiplication, with a perhaps substantial impact on the pulse shape near the peak, which would have a direct impact on the trailing edge.

%
%
\section{Extension to dual-phase detectors}
\label{s:Extension}

Dual-phase detectors, like GPSCs, utilize gas proportional scintillation to measure ionization signals produced in a target volume.  
The target volume, however, is liquid, yielding significantly larger target densities while maintaining the low-energy sensitivity of GPSCs.
Dual-phase proportional scintillation counters are subject to systematic effects beyond the scope of the current work.

Extension of the present work would require inclusion of detector and electrode geometries, electron drift speed and diffusion constants corresponding to the medium and applied electric field, and the effects of the liquid-gas interface which include a finite electron lifetime on this surface and the effects of lateral fields on electron extraction location.
Given the density of liquid as compared to the gas, the energy deposition topology will also be drastically different. 

There are additional systematic effects specific to any given detector, such as material reflectivity, scintillation properties, bulk electron absorption, non-uniform electric fields, and photodetector properties.
The complexity of a dual-phase detector can be considerable, and while the model presented here may allow for reasonable prediction of the pulse shape, it behooves experimenters to modify the model presented here as necessary (see, e.g., Ref~\cite{Mock2014}).

%
%
\section{Conclusions}
\label{s:Conclusions}

A model for description of S2 event shapes from proportional scintillation counters was presented.
The model was based primarily on existing values and measurements from the literature.
The only free parameters in this model were the singlet / triplet ratio in the S2 light production, and the scattering length of the wavelength shifter, tetraphenyl butadiene.

This model was compared to an analytic prediction for a simplified case, and compared to experimental data in a full detailed exploration. 
The discrepancy from the analytic model was negligible, and the full model was in close agreement with the experimental energy spectrum. 
The final pulse shape deviated from the experimental results, with the most likely cause being an oversimplified electric field model.

This work can be used to model the signal from GPSCs, and with extensions, dual-phase proportional scintillation counters.
It can be used to study energy and position reconstruction, scintillation and ionization properties, optical parameters, and the impact of design parameters on future detectors.

We would like to thank M. Szydagis, S. Sangiorgio, and B. Lenardo for helpful discussions. This work was performed under the auspices of the U.S. Department of Energy by Lawrence Livermore National Laboratory under Contract DE-AC52-07NA27344. Funded by Lab-wide LDRD and the Department of Homeland Security ARI. LLNL-JRNL-680202.

%
%
\bibliography{apssamp}

\end{document}